\documentclass[12pt]{article}
\usepackage{times,multicol}
\usepackage{geometry}
\usepackage[utf8]{inputenc}
\usepackage{enumitem,amssymb}
\usepackage{ragged2e}
\newlist{thematic}{itemize}{8}
\setlist[thematic]{label=$\square$}
\usepackage{pifont}

\usepackage[square,sort,numbers]{natbib}
\usepackage{authblk}
\usepackage{aas_macros}
\usepackage{booktabs}
\usepackage{hyperref}
\usepackage{wrapfig}
\usepackage{graphicx,color,rotating}
\usepackage{booktabs}
\usepackage{caption}
\usepackage{lineno,amssymb}
\usepackage[T1]{fontenc}

\begin{document}
\raggedright
\huge
Astro2020 Science White Paper \linebreak

Multi-Messenger Astronomy with Extremely Large Telescopes \linebreak
\normalsize

\noindent \textbf{Thematic Areas:} \hspace*{60pt} $\square$ Planetary Systems \hspace*{10pt} $\square$ Star and Planet Formation \hspace*{20pt}\linebreak
\makebox[0pt][l]{$\square$}\raisebox{.15ex}{\hspace{0.1em}$\checkmark$} Formation and Evolution of Compact Objects \hspace*{31pt} 
\makebox[0pt][l]{$\square$}\raisebox{.15ex}{\hspace{0.1em}$\checkmark$}  Cosmology and Fundamental Physics \linebreak
  $\square$  Stars and Stellar Evolution \hspace*{1pt} $\square$ Resolved Stellar Populations and their Environments \hspace*{40pt} \linebreak
  $\square$    Galaxy Evolution   \hspace*{45pt} \makebox[0pt][l]{$\square$}\raisebox{.15ex}{\hspace{0.1em}$\checkmark$}             Multi-Messenger Astronomy and Astrophysics \hspace*{65pt} \linebreak
  
\textbf{Principal Author:}

Name:	Ryan Chornock
 \linebreak						
Institution:  Ohio University
 \linebreak
Email: chornock@ohio.edu
 \linebreak
Phone: 740-593-1765
 \linebreak
 
\textbf{Co-authors:}
  \linebreak

\vspace{-0.1in}
Philip S. Cowperthwaite (Carnegie); 
Raffaella Margutti (Northwestern); 
Dan Milisavljevic (Purdue);
Kate D. Alexander (Northwestern);
Igor Andreoni (Caltech);
Iair Arcavi (Tel Aviv University);
Adriano Baldeschi (Northwestern);
Jennifer Barnes (Columbia);
Eric Bellm (U Washington);
Paz Beniamini (GWU);
Edo Berger (Harvard);
Christopher P.~L.~Berry (Northwestern);
Federica Bianco (U Delaware);
Peter K. Blanchard (Harvard);
Joshua S. Bloom (UC Berkeley);
Sarah Burke-Spolaor (West Virginia);
Eric Burns (Goddard);
Dario Carbone (Texas Tech University);
S. Bradley Cenko (Goddard);
Deanne Coppejans (Northwestern);
Alessandra Corsi (Texas Tech University);
Michael Coughlin (Caltech);
Maria R. Drout (U Toronto);
Tarraneh Eftekhari (Harvard);
Ryan J.~Foley (UC Santa Cruz);
Wen-fai Fong (Northwestern);
Ori Fox (STScI);
Dale A.~Frail (NRAO);
Dimitrios Giannios (Purdue);
V. Zach Golkhou (Washington);
Sebastian Gomez (Harvard);
Melissa Graham (U Washington);
Or Graur (Harvard);
Aprajita Hajela (Northwestern);
Gregg Hallinan (Caltech);
Chad Hanna (Penn State);
Kenta Hotokezaka (Princeton);
Vicky Kalogera (Northwestern);
Daniel Kasen (UC Berkeley);
Mansi Kasliwal (Caltech);
Adithan Kathirgamaraju (Purdue);
Wolfgang E. Kerzendorf (NYU);
Charles D.~Kilpatrick (UC Santa Cruz);
Tanmoy Laskar (U Bath);
Emily Levesque (U Washington);
Andrew MacFadyen (NYU);
Phillip Macias (UC Santa Cruz);
Ben Margalit (UC Berkeley);
Thomas Matheson (NOAO);
Brian D.~Metzger (Columbia University);
Adam A. Miller (Northwestern);
Maryam Modjaz (NYU);
Kohta Murase (Penn State);
Ariadna Murguia-Berthier (UC Santa Cruz);
Samaya Nissanke (UvA);
Antonella Palmese (Fermilab);
Chris Pankow (Northwestern);
Kerry Paterson (Northwestern);
Locke Patton (Harvard);
Rosalba Perna (Stony Brook);
David Radice (Princeton);
Enrico Ramirez-Ruiz (UC Santa Cruz);
Armin Rest (STScI);
Jeonghee Rho (SETI Institute);
Cesar Rojas-Bravo (UC Santa Cruz);
Nathaniel C. Roth (Maryland);
Mohammad Safarzadeh (Arizona State);
David Sand (U Arizona);
Boris Sbarufatti (Penn State);
Daniel M.~Siegel (Columbia);
Lorenzo Sironi (Columbia);
Marcelle Soares-Santos (Brandeis U);
Niharika Sravan (Purdue);
Sumner Starrfield (Arizona State);
Rachel A.~Street (Las Cumbres Observatory);
Guy S. Stringfellow (University of Colorado Boulder);
Alexander Tchekhovskoy (Northwestern);
Giacomo Terreran (Northwestern);
Stefano Valenti (UC Davis);
V. Ashley Villar (Harvard);
Yihan Wang (Stony Brook);
J. Craig Wheeler (UT Austin);
G. Grant Williams (MMT Observatory);
Jonathan Zrake (Columbia).


\justify

\smallskip
\textbf{Abstract:}

The field of time-domain astrophysics has entered the era of Multi-messenger Astronomy (MMA). One key science goal for the next decade (and beyond) will be to characterize gravitational wave (GW) and neutrino sources using the next generation of Extremely Large Telescopes (ELTs). These studies will have a broad impact across astrophysics, informing our knowledge of the production and enrichment history of the heaviest chemical elements, constrain the dense matter equation of state,  provide independent constraints on cosmology, increase our understanding of particle acceleration in shocks and jets, and study the lives of black holes in the universe. Future GW detectors will greatly improve their sensitivity during the coming decade, as will near-infrared telescopes capable of independently finding kilonovae from neutron star mergers. However, the electromagnetic counterparts to high-frequency (LIGO/Virgo band) GW sources will be distant and faint and thus demand ELT capabilities for characterization. ELTs will be important and necessary contributors to an advanced and complete multi-messenger network. 

\pagebreak


\setlength\parindent{1cm}

\section{Larger Context}
A major frontier in the study of astronomical sources is the development of multi-messenger astronomy (MMA). The field of MMA began with the detection of neutrinos from the Sun and from SN~1987A, but it has recently achieved prominence with the detection of a broad-spectrum electromagnetic (EM) counterpart to the binary neutron star merger (BNS) and gravitational wave (GW) source GW170817 \citep{mmapaper} and the identification of a blazar as the source of the high-energy neutrino event IceCube-170922A \citep{icecube}.  The purpose of this white paper is to argue for the essential role played by Extremely Large Telescopes (ELTs; $\gtrsim$20 m in aperture) in fully realizing the potential of this rapidly emerging field.

We anticipate that the field of MMA will undergo significant development over the next few years (compare with the state of the field for the last decadal survey; \cite{bloom2010}) 
and major advances will occur before the ELTs see first light. Some of the most important developments we can expect by then are: (i) Advanced LIGO/Virgo run O3 and future staged upgrades will occur as the detectors approach design sensitivity around 2022 \citep{abbottlrr}; (ii) additional GW observatories such as KAGRA and LIGO-India will begin regular operations; (iii) the next phase of upgrades to GW sensitivity such as A$+$ will be underway or nearing completion; (iv) next-generation neutrino observatories such as KM3NeT and IceCube-Gen2 will have been completed.  These increases in sensitivity will lead to more detections of counterparts and we will know more basic facts about the populations of sources by the ELT era. However, we expect that major science questions will remain open that can only be addressed with the large apertures of ELTs, which can observe fainter targets and larger, more representative samples than current 8--10~m facilities.

This white 
paper will focus exclusively on the high-frequency ($>$Hz) regime
for GW detections, which covers stellar-mass objects. ELTs will also
contribute critically to supermassive black hole GW science in the
nHz--$\mu$Hz band (as pulsar timing arrays (PTAs) steadily improve their sensitivity), and intermediate-mass GW science in the $\gtrsim$mHz band (following the anticipated launch of LISA). These cases are discussed in more detail in separate white papers (e.g., \cite{kelleyWP}).

\begin{figure}[ht!]
\centering
\includegraphics[angle=0,width=3in]{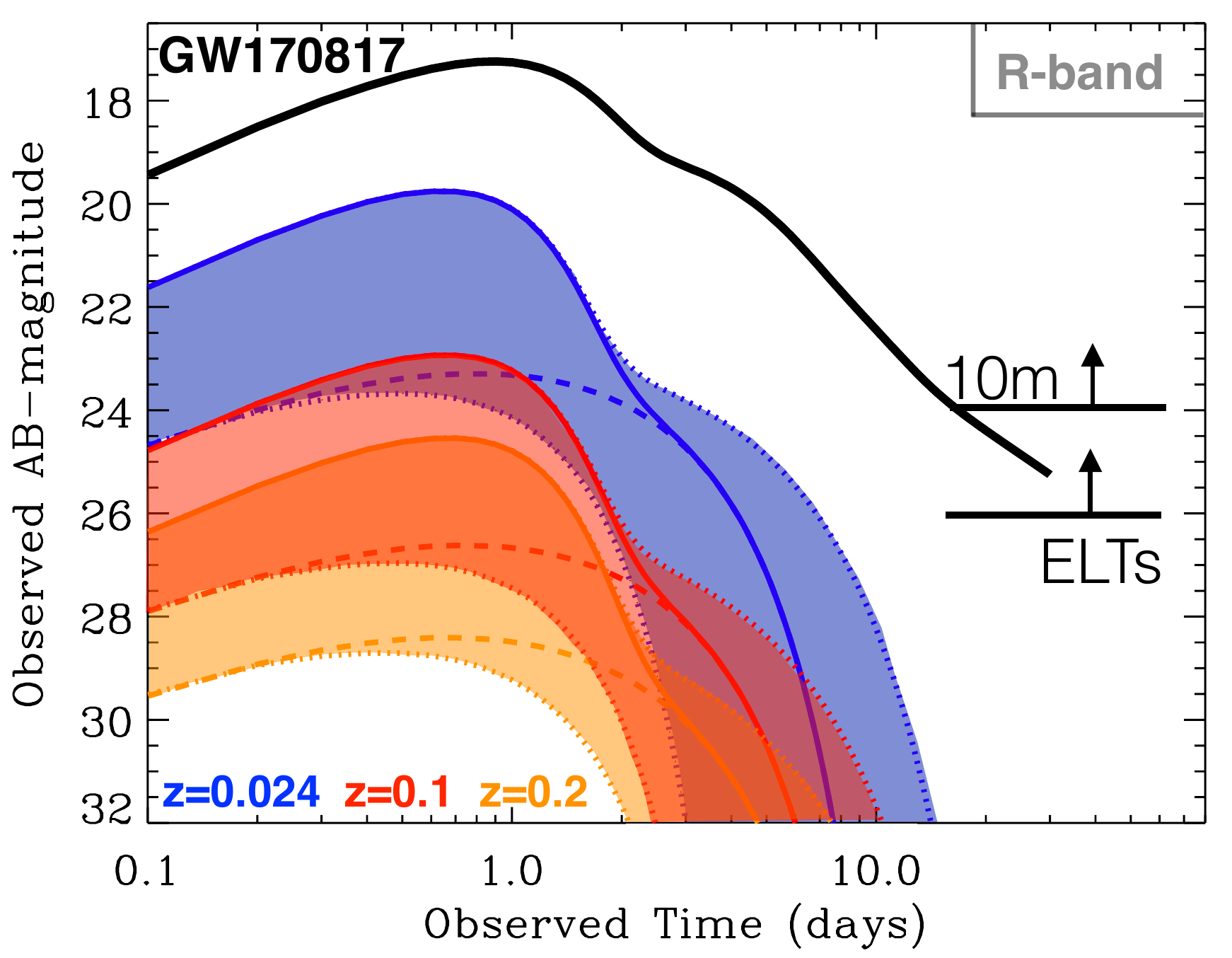}
\includegraphics[angle=0,width=3in]{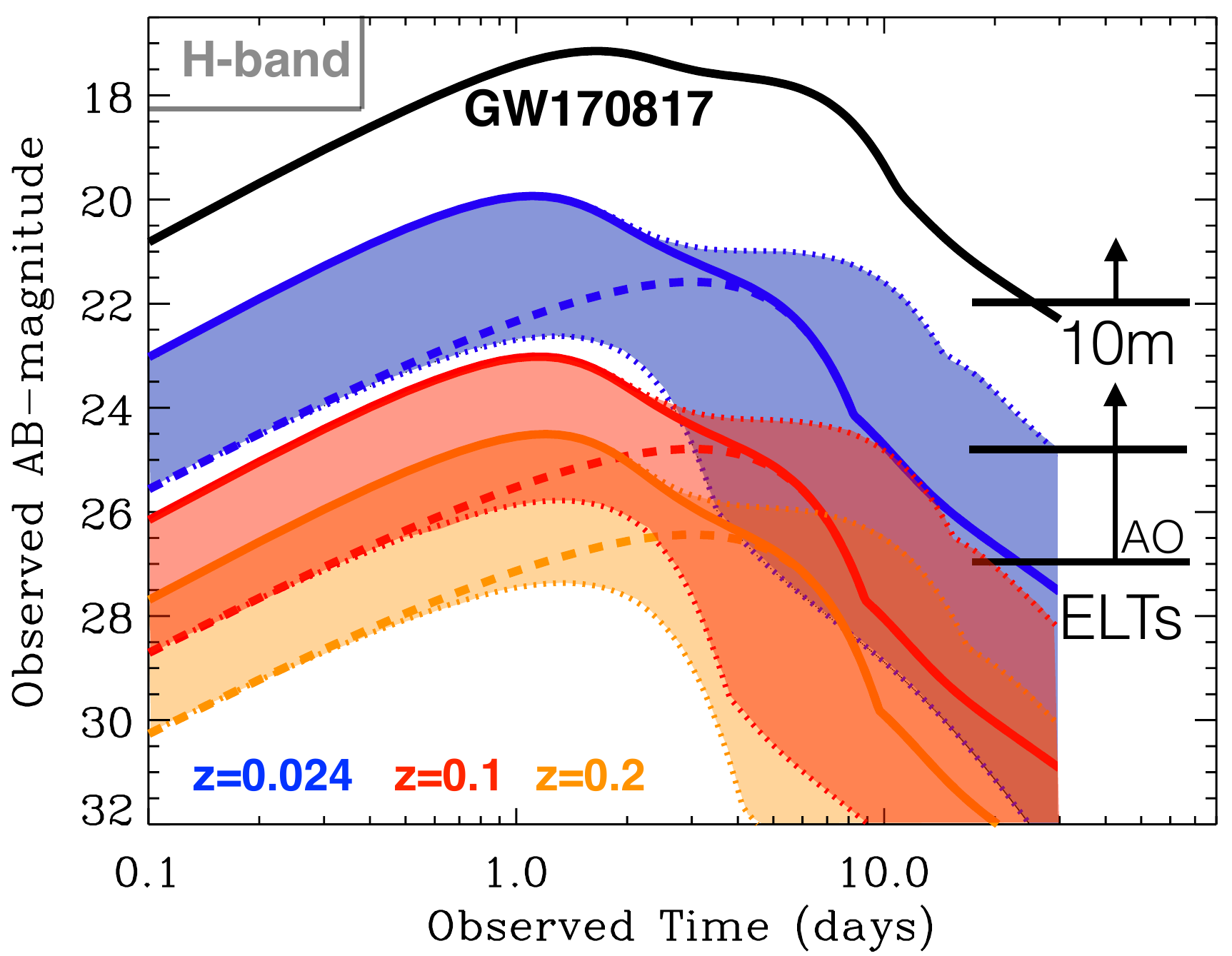}
\caption{Kilonova emission in the optical (R-band, left  panel) and NIR (H-band, right panel) at $z=0.024$ ($d\approx100$ Mpc, blue), $z=0.1$ ($d\approx500$~Mpc; red) and $z=0.2$ (orange), computed following \cite{Villar17}. Fiducial sensitivities for reasonable S/N ratios in an hour-long spectroscopic observation for both current generation 8--10 m telescopes and ELTs are marked, including the improvements in the NIR sensitivity due to adaptive optics (AO).  Black thick line: GW170817 (at 40 Mpc). For each two-component model, the shaded area marks the brightness range corresponding to $M_{ej,red}=0.001-0.05\,\rm{M_{\odot}}$ and $v_{ej,red}=0.1\,\rm{c}$. All of the models include the contribution from a blue kilonova component with similar properties as in GW170817 ($M_{ej,blue}=0.01\,\rm{M_{\odot}}$ and $v_{ej,blue}=0.3\,\rm{c}$). Thick lines: expected total emission for $M_{ej,red}=0.01\,\rm{M_{\odot}}$. Dashed lines: contribution of only the red component with $M_{ej,red}=0.01\,\rm{M_{\odot}}$.  The blue component has a major effect on the optical fluxes, but its origin is obscure. The increased sensitivity of the ELTs in the optical and NIR is necessary to characterize the most distant compact object mergers and to sample the potential diversity of their kilonova emission.}
\label{fig:RH}
  \vspace{-0.1in}
\end{figure}

Despite this uncertainty about the future of a rapidly-evolving field, some aspects of the landscape of the study of high-frequency GW counterparts in the era of ELTs are predictable.  The increased number of functional GW interferometers will substantially increase the localization accuracy of sources and the corresponding ability to find EM counterparts.  However, the increased sensitivity of the observatories will also probe more distant populations of sources. GW interferometers are sensitive to the gravitational strain $h$, which scales as $1/d$ (where $d$ is the distance to a source), while optical telescopes are sensitive to the energy flux, which scales as $1/d^2$.  Although the ultimate design sensitivity of Advanced LIGO will be sensitive to BNS systems out to $\sim$200 Mpc (and BH-NS star mergers at several hundred Mpc), the A$+$ upgrades\footnote{\url{https://dcc.ligo.org/LIGO-T1800042/public}} by the 2025 timeframe will aim to increase the sensitivity by an additional factor of $\sim$2.  Therefore, \textbf{planned technology developments in GW interferometers will rapidly start producing detections at distances where the expected optical counterparts are exceedingly dim, necessitating the collecting area of ELTs to study them in detail} (see Figure~\ref{fig:RH}).  Future deep, wide-area EM surveys, such as with LSST and {\it WFIRST},  will also have synergies with ELTs \cite{mgWP}, including the potential to discover very distant BNS mergers independent of a GW trigger (e.g., \cite{scolnic18}).

The initial detection of an EM counterpart to a BNS merger has demonstrated that at least some short-duration gamma-ray bursts (sGRBs) are connected to BNS mergers and that significant amounts of material enriched by $r$-process nucleosynthesis are ejected in an event known as a kilonova (e.g., \cite{Cowperthwaite17,drout17,gbmbnspaper,kasen2017,kasliwal17,pian17,smartt17,tanvir17}). 
Combining the information gleaned from the EM counterpart with the GW signal can set constraints on fundamental nuclear physics (through the neutron star equation of state) and cosmology (through the use of these sources as standard sirens; \cite{schutz86,hhss}).  This subject is further described in a complementary white paper \cite{gwemWP}.  The future work proposed here will clarify the diversity of outcomes of BNS mergers, as well as potential differences with the outcomes of NS-BH mergers. 

We note that stellar-mass binary black hole systems, despite the intriguing report of a possible associated gamma-ray signal with {\it Fermi} \citep{gbmpaper}, have not yet been conclusively associated with any EM counterpart. In the event of the detection of such counterparts (or of a counterpart to any novel GW source class), the potential discovery space is large. 
ELTs will certainly make major contributions to studies of such novel sources in the likely case that the optical/NIR counterparts are faint.
However, in the absence of reliable predictions for these unknown EM counterparts, we will focus on kilonovae here for specificity.  

\vspace{-0.15in}
\section{Detailed ELT Studies of GW Counterparts} 

The prime counterparts to high-frequency GW sources at optical/near-infrared (NIR) wavelengths are kilonovae, which are produced in compact object mergers containing neutron stars. 
They are the result of the ejection at velocities $\sim$0.1--0.3$c$ of 10$^{-3}$--0.05~M$_{\odot}$ of material enriched in heavy elements by $r$-process nucleosynthesis and powered by the decay of radioactive isotopes \citep{metzger10}.  Theoretical studies of the merger process have identified several physically distinct ejection processes, including tidal tails, squeezed dynamical ejecta, accretion disk winds, and jets. The details are expected to be sensitive to the masses, radii, and tidal deformabilities of the individual neutron stars, which in turn depend on the physics of dense nuclear matter \citep{fm16}.

\begin{figure}[h!]
\centering
  \begin{minipage}[c]{0.65\textwidth}
\includegraphics[width=\textwidth]{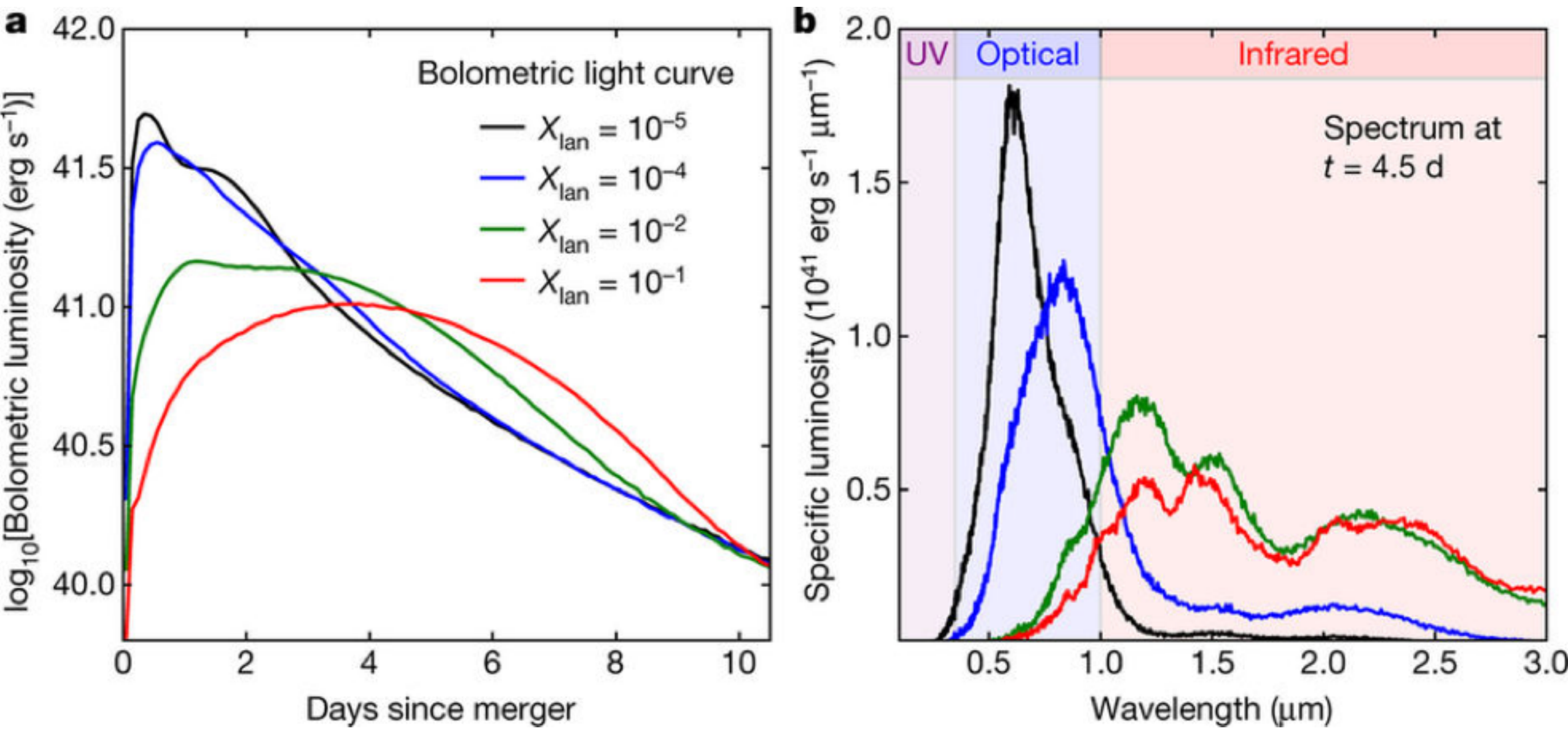}
  \end{minipage}\hfill
  \begin{minipage}[c]{0.32\textwidth}
\caption{As the lanthanide abundance increases in a fiducial kilonova, the  higher opacity results in a longer diffusion timescale, which makes the light curves fainter and broader, while it pushes the flux from the optical to the NIR. The detailed patterns of undulations in the spectra are sensitive to the chemical composition.  From \cite{kasen2017}.}
\label{fig:knspec}
  \end{minipage}
  \vspace{-0.25in}
\end{figure}

If the neutron star debris has a sufficiently low electron fraction, $Y_e<0.23$, neutron captures and $r$-process nucleosynthesis will produce the heaviest elements, including the lanthanide series.  Radiative transfer simulations have shown that the high opacity of lanthanide-rich material has a profound effect on the observed light curves (see Figure~\ref{fig:knspec}).  Material with low lanthanide abundances emits primarily in the optical, while the signatures of high lanthanide abundances are present in the NIR (e.g., \cite{kasen2013,kasen2017,tanakaKN,wollaegerKN}). Therefore, both optical and IR spectroscopy are necessary to be sensitive to the presence of material with a range of compositions.  

A central goal of this future work is to determine the range of outcomes of BNS mergers and the primary drivers of observed diversity.  Simulations have shown that the different ejecta components have properties (masses, velocities, composition) that are sensitive to the properties of the neutron stars.
Studies of sGRBs have already shown that the optical counterparts of presumed compact object mergers exhibit diverse behavior, even after accounting for the afterglow component (e.g., \cite{fong17,gompertz18}).  In addition, we expect that NS-BH systems will differ from BNS mergers.  Due to the stronger GW signal, they will be detected at larger distances than BNS systems on average and thus are expected to have fainter optical counterparts, requiring ELTs.

An unresolved issue after GW170817 is the origin of the blue emission at the earliest times, which was unexpectedly bright.  Despite the neutron-rich nature of debris from neutron stars, "blue" (i.e., lanthanide-poor) kilonova components were predicted to exist and can have several possible origins, including in the dynamical ejecta, accretion disk winds, or after neutrino irradiation by a long-lived hypermassive neutron star.  The presence of these sources for the ejecta associated with the blue component is potentially informative about the properties of the neutron stars (e.g., the radii or the lifetime of a HMNS before collapse to a black hole; \cite{fm16}).
Some authors have proposed additional sources of energy input at the earliest times, such as from free neutron decay or shocks created in the interaction of the jet with the merger debris (e.g., \cite{pk18,gott18}).  Better understanding of these phenomena can therefore shed light on the properties of dense nuclear matter.  Observations at the earliest times, as soon as a counterpart has been identified, will be necessary to constrain these scenarios, requiring rapid-response target of opportunity capabilities with large-aperture facilities.

An additional major source of phenomenological variation in kilonovae is likely to be orientation effects. The most obvious is that the connection with sGRBs demonstrates that some lines of sight will have a relativistic jet pointed directly at the observer.  However, the presence of a well-defined angular momentum axis due to the orbit of the binary means that the ejection of non-relativistic kilonova material during the merger is unlikely to be spherically symmetric.  Understanding the geometry of these events and for the separate components is important to properly measure the ejecta masses, nucleosynthetic outputs, and to accurately constrain the distances for cosmology.  Spectropolarimetry has the potential to directly probe the ejecta geometry \citep{bulla18} and should be performed with ELTs for the nearest and brightest events.
These constraints would be complementary to those of the potentially jetted component studied through non-thermal emission in the radio and X-ray bands.

The large apertures of ELTs will be necessary to obtain late-time observations of kilonovae.  Late-time spectroscopy of kilonovae in the nebular phase has the potential to be highly informative about the velocities and composition of the slowest ejecta, just as for supernovae \cite{danWP}.  Photometric data obtained at still later times can constrain the still highly-uncertain rates of heating and thermalization and search for evidence of individual dominant radioactive species \citep{barnes16}.

The primary desired technical capability besides the large aperture of an ELT is deep spectroscopy with broad wavelength coverage across both the optical and NIR bands.  The expected velocities in GW counterparts are high, so any low-to-moderate spectral resolution is acceptable as long as the sky subtraction is reliable. 
An ideal instrument for this work that could observe the whole spectral window simultaneously might resemble X-Shooter on the VLT or SCORPIO at Gemini-South.  We note that no such instrument is envisioned in the initial instrument suites for GMT or TMT.  Therefore, we encourage further development of proposed instrument concepts such as ARISE at TMT for the second generation suite of instrumentation.

\section{Relationships to Other Facilities}
\vspace{-0.1in}

ELTs cannot perform this science in isolation and will require complementary facilities across the EM spectrum from radio to $\gamma$-rays. Even in optimistic scenarios, GW localizations will remain larger than 10 sq. deg. in the next decade.  ELTs will not have the field of view to efficiently monitor such large sky areas to find potential optical counterparts.  Other wide-field imagers at smaller-aperture facilities will be critical to initially discover the optical counterparts. There will be numerous small telescopes searching for the nearest GW counterparts (e.g., \cite{coulter17,ValentiDLT,arcavi17}), but the more distant ones will require facilities such as LSST (e.g., \cite{marguttiWP}).  ELTs will likely be used primarily for follow-up observations once the counterpart is identified.

Depending on the adopted cadences and survey strategies, deep wide-field surveys with LSST and {\it WFIRST} will have the ability to detect kilonovae independent of GW triggers (e.g., \cite{scolnic18}).  Due to their rarity, these kilonovae will typically be distant and faint, and will require ELTs for spectroscopic observations. 

{\it JWST} will be capable of producing high S/N ratio spectra in the near-IR and mid-IR for a handful of kilonovae that are complementary to the observations discussed here. However, the finite mission lifetime, limited number of ToO interrupts, and sky pointing constraints will limit the sample size that it can observe.  ELTs will be necessary to obtain better-sampled time series of observations for single targets and to build up larger samples of a wider variety of sources.

\vspace{-0.1in}
\section{Goals for Multi-messenger Astronomy with ELTs}
\begin{itemize}
    \item With ELT studies of counterparts to high-frequency GW sources, we can directly observe {\bf the spectral signatures of the heaviest elements} at the sites of their production. A sample of well-observed objects is necessary to determine the dominant factors resulting in the variation in the yields.
This is complementary to studies of the abundance patterns of neutron-capture elements in metal-poor stars in determining the buildup of the periodic table over cosmic time.

\item The relative fractions of material in the different possible ejecta components of kilonovae will constrain the {\bf physics of matter ejection in compact object mergers}.  These processes are interesting in their own right, but are also sensitive to the properties of neutron stars and will help illuminate the dense matter equation of state.  The astrophysical information about the sky location, distance, inclination of the binary, and ejected mass can be combined with the gravitational wave strain data to produce tighter constraints on the properties of the system, such as the tidal deformabilities of neutron stars.

\item Spectroscopic confirmation of distant optical counterparts to GW sources is necessary for {\bf standard siren cosmology}.  While some studies can rely on redshifts from host-galaxy spectroscopy, anything that depends on the properties of the transient, such as estimates of the viewing angle, will require ELTs. At the design sensitivity of Advanced LIGO, BH-NS mergers will already be detectable by GW detectors out to several hundred Mpc, while increases in sensitivity of only $\sim$3 in GW detectors will produce detections of BNS at distances of $\sim$500 Mpc. As shown in Figure~\ref{fig:RH}, at those distances the expected optical counterparts will be too faint for spectroscopy with 10~m class facilities.  After the NSF invests in the A$+$ LIGO upgrades, and if it commits to longer-term plans using the LIGO Voyager or LIGO Cosmic Explorer proposals in the late-2020's or 2030's, ELTs will be necessary to actually perform MMA with many of the GW sources detected by those facilities.

\item The only TeV or PeV candidate neutrino source known to date is a low-redshift blazar (TXS 0506+056; \cite{icecube}), but there is no concordance model \cite{keivani} and the role of ELTs is not clear.  There are other potential {\bf sources of high-energy astrophysical neutrinos} including energetic jets (e.g., gamma-ray bursts and tidal disruption events) and SN shocks with dense media, whose science case is described in the white paper by Ackermann et al. \cite{neutrinoWP}. Rapid follow-up observations using an ELT will be highly desirable to elucidate these processes.  In addition, neutrinos at MeV energies are produced by supernovae, as the detection of SN 1987A demonstrated.  A supernova exploding behind $A_V>25$~mag of extinction in the Galactic plane would produce copious neutrinos and GWs, but could only be studied in the optical/NIR using an ELT.  

\item 
For low-frequency GW sources, ELTs will contribute to \textbf{in-depth MMA studies of supermassive black hole science} and plasma processes in active galactic nuclei, in addition to characterizing host galaxy dynamics using integral-field units to resolve diffraction-limited scales.
For discussion of science cases in those GW regimes, see the related MMA
white papers by Kelley et al. for PTAs \cite{kelleyWP} and Holley-Bockelmann et al. for LISA.

\end{itemize}

\pagebreak
\begin{multicols}{2}
\setlength{\bibsep}{7pt plus 0.3ex}

\end{multicols}

\end{document}